\documentclass{llncs}
\usepackage{graphicx}
\usepackage{float}
\usepackage{cite}

\usepackage{url}
\usepackage[utf8]{inputenc} 
\usepackage{amsmath}
\usepackage{amsfonts}
\usepackage[usenames,dvipsnames]{color}

\usepackage{todonotes}

\begin{document}
%
%

\title{Learning from the News: Predicting Entity Popularity on Twitter}

\author{Pedro Saleiro$^{1,2}$ and Carlos Soares$^{1,3}$}

\institute{DEI-FEUP$^{1}$, LIACC$^{2}$, INESC-TEC$^{3}$,\\University of Porto,\\Rua Dr. Roberto Frias, s/n, Porto, Portugal\\\email{\{pssc,csoares\}@fe.up.pt}
}
\maketitle
\begin{abstract}

In this work, we tackle the problem of predicting entity popularity on Twitter based on the news cycle. We apply a supervised learning approach and extract four types of features: (i) signal, (ii) textual, (iii) sentiment and (iv) semantic, which we use to predict whether the popularity of a given entity will be high or low in the following hours. We run several experiments on six different entities in a dataset of over 150M tweets and 5M news and obtained F1 scores over 0.70.  Error analysis indicates that news perform better on predicting entity popularity on Twitter when they are the primary information source of the event, in opposition to events such as live TV broadcasts, political debates or football matches.
\end{abstract}


\section{Introduction}\label{sec:intro}

Online publication of news articles has become a standard behavior of news outlets, while the public joined the movement either using desktop or mobile terminals. The resulting setup consists of a cooperative dialog between news outlets and the public at large. Latest events are covered and commented by both parties in a continuous basis through the social media, such as Twitter. When sharing or commenting news on social media, users tend to mention the most predominant entities mentioned in the news story \cite{Saleiro2016}. Therefore, entities, such as personalities, organizations, companies or geographic locations, can act as latent interlinks between online news and social media. 

Online Reputation Monitoring (ORM) focuses on continuously tracking what is being said about entities (e.g. politicians) on social media and online news. Automatic collection and processing of comments and opinions on social media is now crucial to understand the reputation of individual personalities and organizations and therefore to manage their public relations. However, ORM systems would be even more useful if they would be able to know in advance if social media users will talk a lot about the target entities or not. For instance, on April 4th 2016, the UK Prime-minister, David Cameron, was mentioned on the news regarding the Panama Papers story. He didn't acknowledge the story in detail on that day. However, the news cycle kept mentioning him about this topic in the following days and his mentions on social media kept very high. He had to publicly address the issue on April 9th, when his reputation had  already been severely damaged, blaming himself for not providing further details earlier. 

We hypothesize that for entities that are frequently mentioned on the news (e.g. politicians) it is possible to establish a predictive link between online news and popularity on social media. We cast the problem as a supervised learning classification approach: to decide whether popularity will be high or low based on features extracted from the news cycle. We define four set of features: signal, textual, sentiment and semantic. We aim to respond to the following research questions: \textbf{RQ1:} Are online news valuable as source of information to effectively predict entity popularity on Twitter? \textbf{RQ2:} Do online news carry different predictive power based on the nature of the entity under study? \textbf{RQ3:} How different thresholds for defining high and low popularity affect the effectiveness of our approach? \textbf{RQ4:} Do performance remains stable for different time of predictions? \textbf{RQ5:} What is the most important feature set for predicting entity popularity on Twitter based on the news cycle? \textbf{RQ6:} Do individual set of features represent different importance for different entities? 

\section{Related Work}

In recent years, a number of research works have studied the relationship and predictive behavior of user response to the publication of online media items, such as, commenting news articles, playing Youtube videos, sharing URLs or retweeting patterns \cite{ bandari2012pulse,yang2011patterns,tsagkias2009predicting,he2014predicting}. The first attempt to predict the volume of user comments for online news articles used both metadata from the news articles and linguistic features \cite{tsagkias2009predicting}. The prediction was divided in two binary classification problems: if an article would get any comments and if it would be high or low number of comments. Similarly other works, found that shallow linguistic features (e.g. TF-IDF or sentiment) and named entities have good predictive power \cite{gottipati2012finding,louis2013makes}. 

Research work more in line with ours, tries to predict the popularity of news articles shares (url sharing) on Twitter based on content features \cite{bandari2012pulse}.  Authors considered the news source, the article's category, the article's author, the subjectivity of the language in the article, and number of named entities in the article as features. Recently, there was a large study of the life cycle of news articles in terms of distribution of visits, tweets and shares over time across different sections of the publisher \cite{castillo2014characterizing}. Their work was able to improve, for some content type, the prediction of web visits using data from social media after ten to twenty minutes of publication.

Other line of work, focused on temporal patterns of user activities and have consistently identified broad classes of temporal patterns based on the presence of a clear peak of activity \cite{crane2008robust,lehmann2012dynamical, romero2011differences, yang2011patterns}. Classes differentiate by the specific amount and duration of activity before and after the peak. Crane and Sornette \cite{crane2008robust} define endogenous or exogenous origin of events based on being triggered by internal aspects of the social network or external, respectively. They find that hashtag popularity is mostly influenced by exogenous factors instead of epidemic spreading. Other work \cite{lehmann2012dynamical} extends these classes by creating distinct clusters of activity based on the distributions in different periods (before, during and after the peak) that can be interpreted based on semantics of hashtags. Consequently, the authors applied text mining techniques to semantically describe hashtag classes. Yang and Leskovec \cite{yang2011patterns} propose a new measure of time series similarity and clustering. Authors obtain six classes of temporal shapes of popularity of a given phrase (meme) associated with a recent event, as well as the ordering of media sources contribution to its popularity.

Recently, Tsytsarau et al. \cite{tsytsarau2014dynamics} studied the time series of news events and their relation to changes of sentiment time series expressed on related topics on social media. Authors proposed a novel framework using time series convolution between the importance of events and media response function, specific to media and event type. Their framework is able to predict time and duration of events as well as shape through time. 

Compared to related work, we focus on a different problem - online reputation monitoring - where it is necessary to track what is being said about an entity on social media on a continuous basis. Therefore, our problem consists on assessing the impact of the news cycle on the time series of popularity of a target entity on social media.

\section{Approach}\label{sec:appr}
The starting point of our hypothesis is that for entities that are frequently mentioned on the news (e.g. politicians) it is possible to predict popularity on social media using signals extracted from the news cycle. The first step towards a solution requires the definition of entity popularity on social media.

\subsection{Entity Popularity}
There are different ways of expressing the notion of popularity on social media. For example, the classical way of defining it is through the number of followers of a Twitter account or likes in a Facebook page. Another notion of popularity, associated with entities, consists on the number of retweets or replies on Twitter and post likes and comments on Facebook.
We define entity popularity based on named entity mentions in social media messages. Mentions consist of specific surface forms of an entity name. For example, ``Cristiano Ronaldo'' might be mentioned also using just ``Ronaldo'' or ``\#CR7''.  

Given an set of entities $E = \{e_1, e_2, ..., e_i, ...\}$, a daily stream of social media messages $S = \{s_1, s_2, ..., s_i, ...\}$ and a daily stream of online news articles $N = \{n_1, n_2, ..., n_i, ...\}$ we are interested in monitoring the mentions of an entity $e_i$ on the social media stream $S$, i.e. the discrete function $f_m(e_i, S)$. Let $T$ be a daily time frame $T = [t_p, t_{p+h}]$, where the time $t_p$ is the time of prediction and $t_{p+h}$ is the prediction horizon time. We want to learn a target popularity function $f_p$ on social media stream $S$ as a function of the given entity $e_i$, the online news stream $N$ and the time frame $T$:
$$f_p(e_i, N, T) = \sum_{t=t_p}^{t=t_{p+h}}  f_m(e_i, S)$$
which corresponds to integrating $f_m(e_i,S)$ over $T$.  

Given a day $d_i$, a time of prediction $t_p$, we extract features from the news stream $N$ until $tp$ and predict $f_p$ until the prediction horizon $tp+h$. We measure popularity on a daily basis, and consequently, we adopted $t_{p+h}$ as 23:59:59 everyday. For example, if $t_p$ equals to 8 a.m, we extract features from $N$ until 07:59:59 and predict $f_p$ in the interval 08:00 - 23:59:59 on day $d_i$. In the case of $t_p$ equals to midnight, we extract features from $N$ on the 24 hours of previous day $d_{i-1}$ to predict $f_p$ for the 24 hours of $d_i$.

We cast the prediction of $f_p(e_i, N, T)$ as a supervised learning classification problem, in which we want to infer the target variable $\hat{f_p}(e_i, N, T) \in \{0,1\}$ defined as:
$$   
 \hat{f_p}=
    \begin{cases}
      0 (\textit{low}), & \text{if}\ P(f_p(e_i, N, T) \leq \delta) = k \\
      1 (\textit{high}), & \text{if}\ P(f_p(e_i, N, T) > \delta) = 1 - k \\
    \end{cases}
$$
where $\delta$ is the inverse of cumulative distribution function at $k$ of $f_p(e_i, N, T)$ as measured in the training set, a similar approach to \cite{tsagkias2009predicting}. For instance, $k=0.5$ corresponds to the median of $f_p(e_i, N, T)$ in the training set and higher values of $k$ mean that $f_p(e_i, N, T)$ has to be higher than $k$ examples on the training set to consider $\hat{f_p}=1$, resulting in a reduced number of training examples of the positive class \textit{high}.

\subsection{News Features}
Previous work has focused on the influence of characteristics of the social media stream $S$ in the adoption and popularity of memes and hashtags \cite{romero2011differences}. In opposition, the main goal of this work is to investigate the predictive power of the online news stream $N$. Therefore we extract four types of features from $N$: (i) \textit{signal}, (ii) \textit{textual}, (iii) \textit{sentiment} and (iv) \textit{semantic}, as depicted in Table \ref{table:feats}. 
One important issue is how can we filter relevant news items to $e_i$. There is no consensus on how to link a news stream $N$ with a social media stream $S$. Some works use URLs from $N$, shared on $S$, to filter simultaneously relevant news articles and social media messages \cite{bandari2012pulse}. As our work is entity oriented, we select news articles with mentions of $e_i$ as our relevant $N$.  

\textit{Signal Features -} This type of features depict the ``signal'' of the news cycle mentioning $e_i$ and we include a set of counting variables as features, focusing on the total number of news mentioning $e_i$ in specific time intervals, mentions on news titles, the average length of news articles, the different number of news outlets that published news mentioning $e_i$ as well as, features specific to the day of the week to capture any seasonal trend on the popularity. The idea is to capture the dynamics of news events, for instance, if $e_i$ has a sudden peak of mentions on $N$, a relevant event might have happened which may influence $f_p$.

\begin{table}[t]
\centering
\caption{Summary of the four type of features we consider: (i) signal, (ii) textual, (iii) sentiment and (iv) semantic, 21 in total}
\hspace*{-0.2cm}\begin{tabular}{llll}
\\
\hline
\textbf{Number} &\textbf{Feature} & \textbf{Description} & \textbf{Type}\\
\hline
\multicolumn{4}{l}{\texttt{Signal}} \\
1 & \textit{news} &   number of news mentions of $e_i$ in $[0, t_{p}]$ in $d_i$ & \textit{Int}\\
2 & \textit{news} $d_{i-1}$ &   number of news mentions of $e_i$ in $[0, t_{p}]$ in $d_{i-1}$ & \textit{Int}\\
3 & \textit{news total} $d_{i-1}$ & number of news mentions of $e_i$ in $[0, 24[$ in $d_{i-1}$ & \textit{Int}\\
4 & \textit{news titles} & number of title mentions in news of $e_i$ in $[0, t_{p}]$ in $d_i$ & \textit{Int}\\
5 & \textit{avg content} & average content length of news of $e_i$ in $[0, t_{p}]$ in $d_i$ & \textit{Float}\\
6 & \textit{sources} & number of different news sources of $e_i$ in $[0, t_{p}]$ in $d_i$ & \textit{Int}\\
7 & \textit{weekday} &   day of week & \textit{Categ}\\
8 & \textit{is weekend} &   true if weekend, false otherwise & \textit{Bool}\\
\hline
\multicolumn{4}{l}{\texttt{Textual}} \\
9-18 & \textit{tfidf titles} &   TF-IDF of news titles $[0, t_{p}]$  in $d_i$& \textit{Float}\\
19-28 & \textit{LDA titles} &   LDA-10 of news titles $[0, t_{p}]$  in $d_i$& \textit{Float}\\
\hline
\multicolumn{4}{l}{\texttt{Sentiment}} \\
29 & \textit{pos} &   number of positive words in news titles $[0, t_{p}]$ in $d_i$& \textit{Int}\\
30 & \textit{neg} &   number of negative words in news titles $[0, t_{p}]$ in $d_i$& \textit{Int}\\
31 & \textit{neu} &   number of neutral words in news titles $[0, t_{p}]$ in $d_i$& \textit{Int}\\
32 & \textit{ratio} &   $positive/negative$ & \textit{Float}\\
33 & \textit{diff} &   $positive - negative$ & \textit{Int}\\
34 & \textit{subjectivity} & $(positive + negative + neutral)/ \sum words$ & \textit{Float}\\
35-44 & \textit{tfidf subj} & TF-IDF of subjective words (pos, neg and neu) & \textit{Float}\\
\hline
\multicolumn{4}{l}{\texttt{Semantic}} \\
45 & \textit{entities} &   number of entities in news $[0, t_{p}]$  in $d_i$ & \textit{Int}\\
46 & \textit{tags} &   number of tags in news $[0, t_{p}]$  in $d_i$ & \textit{Int}\\
47-56 & \textit{tfidf entities} &   TF-IDF of entities in news $[0, t_{p}]$  in $d_i$ & \textit{Float}\\
57-66 & \textit{tfidf tags} &   TF-IDF of news tags $[0, t_{p}]$  in $d_i$ & \textit{Float}\\
\hline
\end{tabular}
\label{table:feats}
\end{table}

\textit{Textual features -} To collect textual features we build a daily profile of the news cycle by aggregating all titles of online news articles mentioning $e_i$ for the daily time frame $[0, t_p]$ in $d_i$. We select the top 10000 most frequent terms (unigrams and bi-grams) in the training set and create a document-term matrix $R$. Two distinct methods were applied to capture textual features. 

The first method is to apply TF-IDF weighting to $R$. We employ singular value decompositiong (SVD) to capture similarity between terms and reduce dimensionality. It computes a low-dimensional linear approximation $\sigma$. The final set of features for training and testing is the  TF-IDF weighted term-document matrix $R$ combined with $\sigma R$ which produces 10 real valued latent features. When testing, the system uses the same 10000 terms from the training data and calculates TF-IDF using the IDF from the training data, as well as, $\sigma$ for applying SVD on test data.

The second method consists in applying Latent Dirichlet allocation (LDA) to generate a topic model of 10 topics (features). The system learns a topic-document distribution $\theta$ and a word distribution over topics $\varphi$ using the training data for a given entity $e_i$. When testing, the system extracts the word distribution of the news title vector $r$ on a test day $d'_i$. Then, by using $\varphi$ learned on training data, it calculates the probability of $r$ belonging to one of the 10 topics learned before. 
The objective of extracting this set of features is to create a characterization of the news stream that mentions $e_i$, namely, which are the most salient terms and phrases on each day $d_i$ as well as the latent topics associated with $e_i$. By learning our classifier we hope to obtain correlations between certain terms and topics and $f_p$.

\textit{Sentiment features -} We include several types of word level sentiment features. The assumption here is that subjective words on the news will result in more reactions on social media, as exposed in \cite{dos2015breaking}.  Once again we extract features from the titles of news mentioning $e_i$ for the daily time frame $[0, t_p]$. We use a sentiment lexicon as \textit{SentiWordNet} to extract subjective terms from the titles daily profile and label them as positive, neutral or negative polarity. We compute count features for number of positive, negative, neutral terms as well as difference and ratio of positive and negatives terms. Similar to textual features we create a TFIDF weighted term-document matrix $R$ using the subjective terms from the title and apply SVD to compute 10 real valued sentiment latent features.

\textit{Semantic features -} We use the number of different named entities recognized in $N$ on day $d_i$  until $tp$, as well as, the number of distinct news category tags extracted from the news feeds metadata. These tags, common in news articles, consist of author annotated terms and phrases that describe a sort of semantic hierarchy of news categories, topics and news stories (e.g. ``european debt crisis''). We create  a TF-IDF weighted entity-document and TF-IDF tag-document matrices and applied SVD to each of them to reduce dimensionality to 10. The idea is to capture interesting entity co-occurrences as well as, news stories that are less transient in time and might be able to trigger popularity on Twitter.

\subsection{Learning Framework}

Let $x$ be the feature vector extracted from the online news stream $N$ on day $d_i$ until $tp$. We want to learn the probability $P(\hat{f_p}=1|X=x)$. This can be done using the inner product between $x$ and a weighting parameter vector $w \in \mathbb{R}$, $\textbf{w}^\top \textbf{x}$. 

Using logistic regression and for binary classification one can unify the definition of $p(\hat{f_p}=1|x)$ and $p(\hat{f_p}=0|x)$ with

$$ p(\hat{f_p}|x) = \frac{1}{1 + e^{-\hat{f_p} w^\top x}}  $$

Given a set of $z$ instance-label pairs ($x_i$,$\hat{f_p}_i$), with $i = 1,...,z$ and $\hat{f_p}_i \in \left\{0,1\right\}$ we solve the binary class L2 penalized logistic regression optimization problem, where $C > 0$  

$$\underset{w}{min\,} \frac{1}{2} w^\top w + C \sum_{i=1}^n \log(1 + e^{- \hat{f_p}_i w^\top x_i})$$

We apply this approach following an entity specific basis, i.e. we train an individual model for each entity. Given a set of entities $E$ to which we want to apply our approach and a training set of example days $D = \{d_1, d_2, ..., d_i, ...\}$, we extract a feature vector $x_i$ for each entity $e_i$ on each training day $d_i$. Therefore, we are able to learn a model of $w$ for each $e_i$. The assumption is that popularity on social media $f_p$ is dependent of the entity $e_i$ and consequently we extract entity specific features from the news stream $N$. For instance, the top 10000 words of the news titles mentioning $e_i$ are not the same for $e_j$. 

\section{Experimental Setup}
\begin{figure}[h]
\centering
    \includegraphics[width=0.8\textwidth]{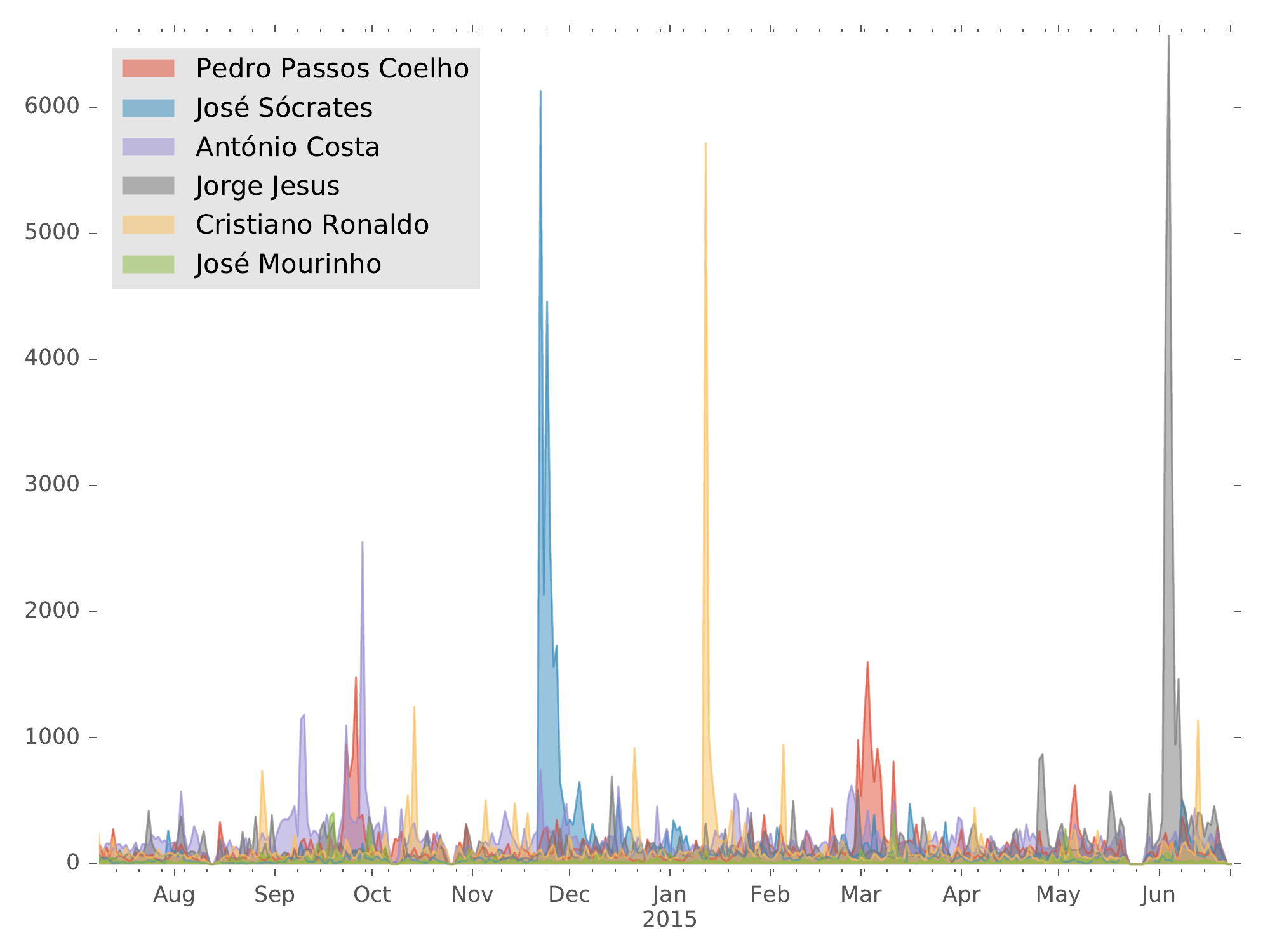}
\caption{Daily popularity on Twitter of entities under study.}
\label{fig:architecture}
\end{figure}

This work uses Portuguese news feeds and tweets collected from January 1, 2013 to January 1, 2016, consisting of over 150 million tweets and 5 million online news articles\footnote{Dataset is available for research purposes. Access requests via e-mail.}. To collect and process raw Twitter data, we use a crawler, which recognizes and disambiguate named entities on Twitter \cite{boanjak2012twitterecho,saleiro2013popstar,saleiro2015popmine}.
News data is provided by a Portuguese online news aggregator. This service handles online news from over 60 Portuguese news outlets and it is able to recognize entities mentioned on the news.

We choose the two most common news categories: politics and football and select the 3 entities with highest number of mentions on the news for both categories. The politicians are two former Prime-ministers, Jos{\'e} S{\'o}crates and Pedro Passos Coelho and the incumbent, Ant{\'o}nio Costa. The football entities are two coaches, Jorge Jesus and Jos{\'e} Mourinho, and the most famous Portuguese football player, Cristiano Ronaldo. 

Figure \ref{fig:architecture} depicts the behavior of daily popularity of the six entities on the selected community stream of Twitter users for  each day from July 2014 until July 2015. As expected, it is easily observable that in some days the popularity on Twitter exhibits bursty patterns. For instance, when Jos{\'e} S{\'o}crates was arrested in November 21st 2014 or when Cristiano Ronaldo won the FIFA Ballon d'Or in January 12th 2015.

\begin{figure}[H]
\centering
    \includegraphics[width=0.99\textwidth]{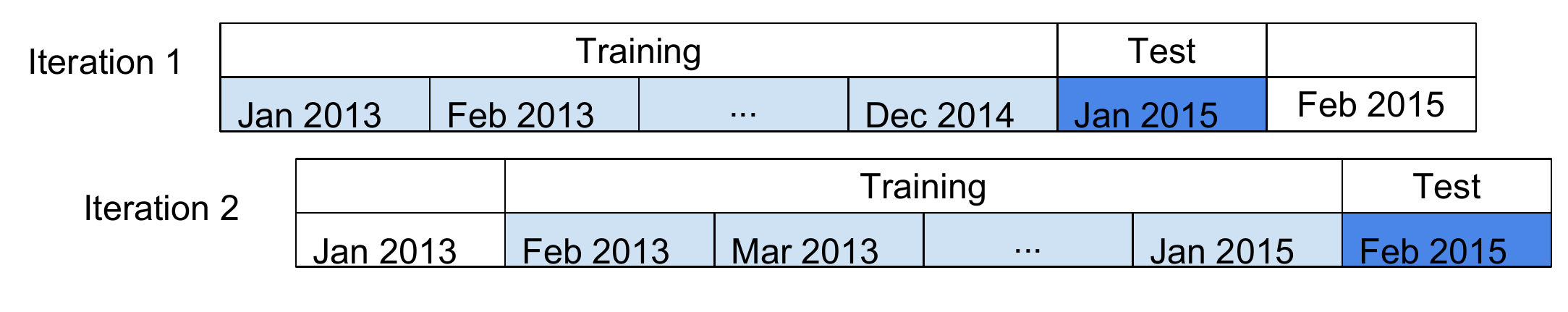}
\caption{Training and testing sliding window - first 2 iterations.}
\label{fig:sliding}
\end{figure}

We defined the years of 2013 and 2014 as training set and the whole year of 2015 as test set. We applied a monthly sliding window setting in which we start by predicting entity popularity for every day of January 2015 (i.e. the test set) using a model trained on the previous 24 months, 730 days (i.e. the training set). Then, we use February 2015 as the test set, using a new model trained on the previous 24 months. Then March and so on, as depicted in Figure \ref{fig:sliding}. We perform this evaluation process, rolling the training and test set until December 2015, resulting in 365 days under evaluation.

The process is applied for each one of the six entities, for different time of predictions $t_p$ and for different values of the decision boundary $k$. We test $tp = {0, 4, 8, 12, 16, 20}$ and $k = {0.5, 0.65, 0.8}$. Therefore, we report results in Section \ref{sec:results} for 18 different experimental settings, for each one of the six entities. The goal is to understand how useful the news cycle is for predicting entity popularity on Twitter for different entities, at different hours of the 24 hours cycle and with different thresholds for considering  popularity as \textit{high} or \textit{low}.

\section{Results and Outlook}\label{sec:results}
Results are depicted in Table \ref{table:res}. We report F1 on positive class since in online reputation monitoring is more valuable to be able to predict \textit{high} popularity than \textit{low}. Nevertheless, we also calculated overall Accuracy results, which were  better than the F1 reported here. Consequently, this means that our system is fairly capable of predicting \textit{low} popularity.
We organize this section based on the research questions we presented in the Introduction (Section 1).
\begin{table}[H]
\centering
\caption{F1 score of popularity \textit{high} as function of $t_p$ and $k$ equal to 0.5, 0.65 and 0.8 respectively.}
\setlength{\tabcolsep}{10pt}
\begin{tabular}{l c c c c c c }
\\
\hline
 \textbf{Entity  \textbackslash       $t_p$(hour)}&\textbf{0} &\textbf{4} & \textbf{8} & \textbf{12} & \textbf{16}& \textbf{20}\\
\hline
 \multicolumn{7}{l}{\textbf{$k=0.50$}} \\
Ant{\'o}nio Costa	& 0,76	& 0,67	& 0,74	& 0,77	& 0,75	& 0,72 \\
Jos{\'e} S{\'o}crates & 	0,77 & 	0,66 & 	0,73 & 	0,75 & 	0,75 & 	0,75 \\
Pedro Passos Coelho &	0,72 & 	0,63 & 	0,70 & 	0,70	 & 0,74	 & 0,71\\
Cristiano Ronaldo	& 0,35	& 0,41 &	0,45 &	0,37	& 0,35& 0,32 \\
Jorge Jesus & 0,73	 & 0,68	 & 0,69	 & 0,68 & 	0,69 & 	0,70 \\
Jos{\'e} Mourinho & 	0,62 & 	0,46 & 	0,51 & 	0,56	 & 0,55	 & 0,45 \\
\hline
\multicolumn{7}{l}{\textbf{$k=0.65$}} \\
Ant{\'o}nio Costa	 & 0,61 & 	0,60 & 	0,66	 & 0,64	 & 0,60 & 	0,60 \\
Jos{\'e} S{\'o}crates	 & 0,63 & 	0,57	 & 0,62 & 	0,66 & 	0,64	 & 0,62 \\
Pedro Passos Coelho	 & 0,58	 & 0,57 & 	0,65	 & 0,67 & 	0,67 & 	0,65 \\
Cristiano Ronaldo & 	0,29 & 	0,35	 & 0,42 & 	0,41 & 	0,36 & 	0,30 \\
Jorge Jesus	 & 0,63	 & 0,61	 & 0,63 & 	0,59	 & 0,62	 & 0,64 \\
Jos{\'e} Mourinho	 & 0,56	 & 0,39	 & 0,48 & 	0,56	 & 0,47	 & 0,38 \\
\hline
\multicolumn{7}{l}{\textbf{$k=0.80$}} \\
Ant{\'o}nio Costa	 & 0,48	 & 0,51 & 	0,55	 & 0,53	 & 0,44 & 	0,49 \\
Jos{\'e} S{\'o}crates	 & 0,48	 & 0,42	 & 0,47	 & 0,53	 & 0,47 & 	0,35 \\
Pedro Passos Coelho & 	0,47	 & 0,46 & 	0,56	 & 0,56 & 	0,52	 & 0,54 \\
Cristiano Ronaldo & 	0,14 & 	0,29	 & 0,31 & 	0,26 & 	0,20	 & 0,21 \\
Jorge Jesus	 & 0,50	 & 0,48 & 	0,51	 & 0,48 & 	0,57	 & 0,56 \\
Jos{\'e} Mourinho & 	0,32 & 	0,32 & 	0,36	 & 0,41	 & 0,41 & 	0,36 \\
\hline
\end{tabular}
\label{table:res}
\end{table}
\textbf{RQ1} and \textbf{RQ2:} Results show that performance varies with each target entity $e_i$. In general, results are better in the case of predicting popularity of politicians. In the case of football personalities, Jorge Jesus exhibits similar results with the three politicians but Jos{\'e} Mourinho and specially Cristiano Ronaldo represent the worst results in our setting. For instance, when Cristiano Ronaldo scores three goals in a match, the burst on popularity is almost immediate and not possible to predict in advance.  

Further analysis showed that online news failed to be informative of popularity in the case of live events covered by other media, such as TV. Interviews and debates on one hand, and live football games on the other, consist of events with unpredictable effects on popularity. Cristiano Ronaldo can be considered a special case in our experiments. He is by far the most famous entity in our experiments and in addition, he is also an active Twitter user with more than 40M followers. This work focus on assessing the predictive power of online news and its limitations. We assume that for Cristiano Ronaldo, endogenous features from the Twitter itself would be necessary to obtain better results. 

\textbf{RQ3:} Our system exhibits top performance with $k=0.5$, which corresponds to balanced training sets, with the same number of $high$ and $low$ popularity examples on each training set. 
Political entities exhibit F1 scores above 0.70 with $k=0.5$. On the other hand, as we increase $k$, performance deteriorates. We observe that for $k=0.8$, the system predicts a very high number of false positives. It is very difficult to predict extreme values of popularity on social media before they happen. We plan to tackle this problem in the future by also including features about the target variable in the current and previous hours, i.e., time-series auto-regressive components.
\begin{figure}[H]
\centering
    \includegraphics[width=0.9\textwidth]{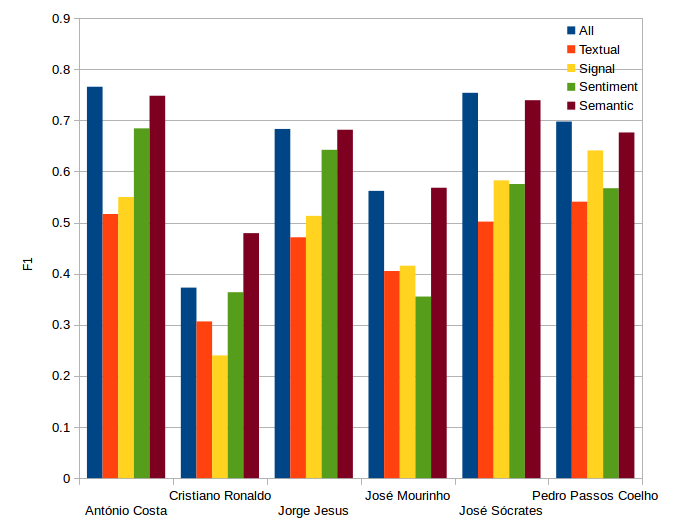}
\caption{Individual feature type F1 score for $t_p=12$ at $k=0.5$.}
\label{fig:feats}
\end{figure}
\textbf{RQ4:} Results show that time of prediction affects the performance of the system, specially for the political entities. In their case, F1 is higher when time of prediction is noon and 4 p.m. which is an evidence that in politics, most of the news events that trigger popularity on social media are broadcast by news outlets in the morning. It is very interesting to compare results for midnight and 4 a.m./8 a.m. The former use the news articles from the previous day, as explained in Section \ref{sec:appr}, while the latter use news articles from the first 4/8 hours of the day under prediction. In some examples, Twitter popularity was triggered by events depicted on the news from the previous day and not from the current day.

\textbf{RQ5} and \textbf{RQ6:}  Figure \ref{fig:feats} tries to answer these two questions. The first observation is that the combination of all groups of features does not lead to substantial improvements. Semantic features alone achieve almost the same F1 score as the combination of all features. However in the case of Mourinho and Ronaldo, the combination of all features lead to worse F1 results than the semantic set alone. 

Sentiment features are the second most important for all entities except Jos{\'e} Mourinho. Signal and Textual features are less important and this was somehow a surprise. Signal features represent the surface behavior of news articles, such as the volume of news mentions of $e_i$ before $t_p$ and we were expecting an higher importance. Regarding Textual features, we believe that news articles often refer to terms and phrases that explain past events in order to contextualize a news article. 

In future work, we consider alternative approaches for predicting future popularity of entities that do not occur everyday on the news, but do have social media public accounts, such as musicians or actors. In opposition, entities that occur often on the news, such as economics ministers and the like, but do not often occur in the social media pose also a different problem.

%
\bibliographystyle{unsrt}
\bibliography{refs}  
%
%
\end{document}